\documentstyle[osa,twocolumn,prl,aps,graphicx]{revtex}
\begin{document}

%\twocolumn[\hsize\textwidth\columnwidth\hsize\csname %
%@twocolumnfalse\endcsname

\wideabs{

\title{Spin-lattice relaxation in the mixed state of 
       YBa$_2$Cu$_3$O$_{7-\delta}$:  \\
       Can we see Doppler-shifted d-wave quasiparticles?}

\author{R. Wortis, A.J. Berlinsky, C. Kallin}

\address{
Department of Physics and Astronomy,
McMaster University, 
1280 Main Street West,
Hamilton, Ontario, Canada L8S-4M1}

\date{\today}

\maketitle

\begin{abstract}

We present calculations of the rate of planar Cu spin lattice
relaxation in the mixed state of YBa$_2$Cu$_3$O$_{7-\delta}$ due to
(i) vortex vibrations and (ii) electron spin-flip scattering.
We emphasize that both mechanisms give position dependent rates due to
the presence of vortices, and hence the magnetization recovery is
characterized by a distribution of rates.
We conclude that relaxation by vortex vibrations is too slow to be a
significant factor in this material.
Using a semiclassical model of Doppler shifted d-wave quasiparticles 
with a linear dispersion around the nodes, our calculation of the
relaxation rate from electron spin-flip scattering shows partial
agreement with experiment.

\end{abstract}


}

\section{Introduction}
\label{intr}

The spin-lattice relaxation rate is a classic probe of the
superconducting state.  
The application of magnetic fields is central to NMR, and
to study spin relaxation in the Meissner state, great ingenuity is
required to make measurements using a field while insuring that the
relaxation takes place in the absence of field.\cite{hs:59}
However, in type-II materials, magnetic field enters the
superconductor naturally in the form of vortices.
The cuprate superconductors are extreme type-II materials in which the
internal field is relatively uniform.

The Slichter group has made measurements of nuclear
spin-lattice relaxation in the vortex state of
YBa$_2$Cu$_3$O$_{7-\delta}$ (YBCO).\cite{data}  
A key feature of the data they obtain is a broad distribution of
relaxation rates.
A distribution of relaxation rates necessarily implies spatial
nonuniformity.
(Because the relaxation rates due to two or more mechanisms at a
given site simply add, if there are multiple mechanisms that are
all uniform, the final result is also uniform.)
In the mixed state of a superconductor there is intrinsic spatial
nonuniformity, namely the vortices.

What mechanisms of relaxation would reflect this nonuniformity?
Possible relaxation mechanisms can be divided into two classes
according to whether they involve an interaction with the magnetic or
the electric quadrupole moment of the nucleus.  
Given the magnetic nature of vortices, it seems reasonable to
focus on the magnetic mechanisms.  
It is possible that vortices induce strains in the crystal
lattice; however, any resulting variation in relaxation by phonons
would be a second order effect.

Perhaps the most direct magnetic mechanism is the oscillation of
magnetic field caused by vortex motions.
Given the small coherence length and the consequent importance of
thermal fluctuations in the cuprates, this mechanism deserves
consideration. 
However, the magnetic mechanism which dominates relaxation in most
conducting materials is electron spin-flip scattering: an electron
exchanges spin with a nucleus and simultaneously scatters to a new
momentum state in such a way as to conserve energy.
In a superconductor with nodes in its gap,
this mechanism is influenced by vortices because the initial and final
states available to the electron depend on the supercurrent velocity
\cite{gev:93}
which is a function of distance from the vortex cores.
In addition, there may be electronic states associated with the cores
themselves. 
In this paper we evaluate the distribution of relaxation arising from   
vortex vibrations as well as that from electron spin-flip scattering. 

Most previous studies\cite{bksv:93,xc:94,vvd:94} of relaxation by
vortex vibrations have focused on the spatial average of the rate. 
We find that this mechanism is in fact strongly position dependent.
However, in the relatively three dimensional compound YBCO, on which
we have focused for comparison with the available experiments, even
the fastest rates we calculate from vortex vibrations are much slower
than what is observed experimentally.  

It is the latter mechanism, electron spin-flip scattering, which
offers the most exciting possibilities as a window on the electronic
states in the superconducting state of the cuprates.
It is now widely accepted that the symmetry of the superconducting
order parameter in YBCO (and many other hole doped cuprates) is
d-wave. 
However, there is less consensus on whether the electronic excitations
of the superconducting state are described by the conventional BCS
picture of weakly interacting quasiparticles.
Microwave conductivity measurements\cite{ah:98} find that the real
part of the conductivity has a Drude-like form with a width which
drops sharply below $T_c$, corresponding to increasing quasiparticle
lifetimes entering the superconducting state.  
However, it is not clear how to reconcile the temperature independent
scattering rate at low temperatures with any known form of scattering.
In addition, it has been argued that recent photoemission measurements
are inconsistent with weakly interacting quasiparticles.\cite{arpes}

It was first pointed out by Volovik\cite{gev:93} that the Doppler
shifting of quasiparticle energies in the presence of supercurrent
flow would generate a nonzero DOS at the Fermi surface in a
superconductor with nodes in the gap.  
If the gap rises linearly near the nodes, the DOS at the Fermi surface
will be proportional to the superfluid velocity.
The supercurrent circulating around a vortex falls off roughly as one
over the distance from the core, but is cut off by the spacing between
vortices. 
Hence, Volovik argued that the DOS at the Fermi surface would be
proportional to one over the spacing between vortices and therefore to
the square root of the applied field.
There will also be a contribution to the density of states from
states associated with the vortex cores, but this is assumed to be
negligible well away from the cores.

Specific heat provides a direct measure of the DOS.
The first experiments were done by Moler, {\it et al},\cite{kam:94}
and since then additional measurements have been made by other
groups\cite{rgjew:98}, 
providing data in a wide range of temperatures and fields. 
Theory\cite{kv:96,sl:97i,vk:97,sl:97ii}
suggests that there should be two scaling regimes:
At low temperatures and high fields, the dominant effect is the
Doppler shifting of quasiparticles, and hence the $\sqrt{H}$ dependence
first predicted by Volovik is expected.
However, at high temperatures and low fields, occupation of
quasiparticle states by Doppler shifting is only significant near the
vortex cores, and elsewhere thermally excited quasiparticles dominate.
In this regime the field dependence of the specific heat is expected
to be linear. 
Much of the data fall in a crossover region between these two regimes. 
Impurities give a nonzero DOS at the Fermi surface even in zero field,
and therefore modify these scaling arguments.\cite{kh:98i,kh:98iii}

Specific heat measurements, however, have associated complications,
including the nontrivial process of separating the electronic
contributions from those of the lattice and of impurities.
Moreover, it has been argued\cite{apr:96,jes:99} 
that the field dependence of the specific
heat observed in the cuprates is not due to their d-wave nature but
rather to variation of the vortex core size independent of the gap
symmetry. 
Therefore, it is beneficial to look for corroboration from other
probes of the DOS. 
A substantial body of work on thermal conductivity is being 
developed from this perspective.\cite{mc:99,kh:98ii,mf:99,vh:99}

Here we study these issues from the perspective of NMR.
Is electron spin-flip scattering the dominant (or at least a
significant) relaxation mechanism in superconducting YBCO?
Are the electronic states well described in terms of Doppler-shifted
d-wave quasiparticles?
To address these questions, we have constructed a model of the
relaxation based on Volovik's semi-classical picture, focusing on
states near the gap nodes.
At experimentally accessible fields, the vortex cores account for
at most a few percent of the total volume of the sample, and
we therefore do not consider here the effect of electronic states
associated specifically with the cores.  
To facilitate comparison with the available experiments, we have
focused on the copper (spin ${3 \over 2}$) nuclei in the Cu-O planes. 

Our results from electron spin-flip scattering are indeed spatially
nonuniform and similar to what is seen experimentally.
However, the correspondence is not entirely satisfactory.  
The distribution of relaxation rates obtained is not as broad as that
seen in experiments nor is there as much weight at the highest rates.
Furthermore, our calculations do not show as strong a field dependence
as is observed.
A more detailed discussion of these issues appears in the conclusion.

This paper proceeds as follows:
Section (\ref{pofw}) defines the distribution of relaxation rates and
the  associated magnetization recovery.
In section (\ref{vv}) we outline our calculation of the rate of
relaxation by vortex vibrations and describe the results of our
numerical evaluation. 
In section (\ref{esf}) we present our calculation of the rate of
relaxation by electron spin-flip scattering and discuss its
evaluation.
Finally, section (\ref{concl}) presents our conclusions.

\section{The magnetization recovery from the rate distribution}
\label{pofw}

The magnetization of the whole sample as a function of time is given
by the integral over all possible relaxation rates, $W_1$, of the 
product of (i) the fraction of spins which relax at a given rate,
$P(W_1)$, and (ii) the magnetization of those spins as a function of
time, $M(W_1,t)$:
\begin{eqnarray}
M(t)
&=&
\int_0^{\infty} P(W_1) M(W_1,t) dW_1
\label{moft}
\end{eqnarray}
We assume that each set of spins characterized by a given $W_1$ is
described by a spin temperature.  
Considering, for example, an upper satellite 
($-{1 \over 2}$ to $-{3 \over 2}$) inversion recovery,
the magnetization of each set of spins will display the following
three exponential decay: \cite{jm:93}
\begin{eqnarray}
M(W_1,t)
&=&
M(W_1,\infty)
\left[ 1 - 2 F {\cal E}(W_1 t) \right] \quad {\rm where}
\nonumber \\
{\cal E}(W_1 t) 
&=&
\left( {1 \over 10} e^{- 2 W_1 t} 
       + {5 \over 10} e^{- 6 W_1 t}
       + {4 \over 10} e^{- 12 W_1 t}
\right). 
\label{mofw1t}
\end{eqnarray}
%
% This form assumes
% 1. a spin temperature
% 2. nuclear Zeeman splitting << thermal energy
% 3. magnetic transition mechanism so $\Delta m = \pm 1$
%
% The derivation is given in my thesis, Appendix B.4, p. 124 ff.
% On the definition of $W_1$, see also /calcs/W1insl.tex.
%
$M(W_1,\infty)$ is the equilibrium magnetization of the transition,
and $F$ is the fraction of spins which are initially inverted
(one ideally, but less in real experiments).
$W_1$, or equivalently $1/T_1$, is defined here to be equal to one
third the rate at which a nuclear spin flips between the states 
$m=-{3 \over 2}$ and $m=-{1 \over 2}$.   
In principle the up and down transition rates are different because
the populations of the two states are different.
However, in practice, the thermal energy is so much larger than the
nuclear Zeeman energy that the population difference is very small and
need only be kept to first order.

In a uniform system, $P(W_1)$ would be a $\delta$-function.
However, in our system the rate of relaxation, whether from electron
spin-flip scattering or from vortex vibrations, depends on position in
the vortex lattice. 
$P(W_1)$ is obtained by adding contributions from all positions:
\begin{eqnarray}
P(W_1)
&=&
{B \over \Phi_o}
\int d{\bf r} \ 
\delta \left( W_1 - W({\bf r}) \right)
\label{pofw1}
\end{eqnarray}
$B$ is the average field {\it in} the sample,
and $\Phi_o$ is the flux quantum. 
Our task is therefore to calculate $W({\bf r})$,
the rate of relaxation at ${\bf r}$.

\section{Relaxation by vortex vibrations}
\label{vv}

\subsection{Calculating $W^{vv}$}
\label{vvcalc}

First we calculate the rate of relaxation due to vortex vibrations,
$W^{vv}$.
We assume the vortices make only small deviations from a perfect
hexagonal lattice,  
and we consider only the case in which the external field is
applied parallel to the c axis.
Furthermore, we assume that the vortex dynamics are overdamped.
Because we focus on YBa$_{2}$Cu$_{3}$O$_{7-\delta}$, which is among
the least strongly layered cuprates, we use a three dimensional
formalism with c-axis anisotropy, neglecting ab anisotropy. 

The relaxation rate due to vortex vibrations, $W^{vv}$, is
proportional to the correlation function of the transverse field:
\cite{bksv:93}
\begin{eqnarray}
\label{w1}
W^{vv}({\bf r})
&=&
{1 \over 2} \gamma_n^2 K({\bf r},\omega=\gamma_n B)
\\
K({\bf r},\omega)
&=&
\int_{-\infty}^{+\infty} dt \ e^{i \omega t}
\langle {\bf h}_{\perp}({\bf r},t) 
        \cdot {\bf h}_{\perp}({\bf r},0) \rangle
\label{krom}
\end{eqnarray}
where ${\bf h}_{\perp}$ is the component of the local magnetic field
which is perpendicular to the c axis,
and where $\langle \rangle$ denotes a thermal average.

The magnetic field is defined by
\begin{eqnarray}
\label{3aheq}
{\bf h}+{\vec \nabla}\times
[\underline{\Lambda}\cdot({\vec \nabla}\times{\bf h})]
&=&
\Phi_o \sum_i \int d {\bf r}_i \ 
\delta_{(3)}({\bf r}-{\bf r}_i)
\nonumber \\
{\rm and} \quad \vec \nabla \cdot {\bf h} &=& 0
\end{eqnarray}
where $\underline{\Lambda}$ is diagonal and
\begin{eqnarray}
\label{3ala}
\Lambda_{xx}=\Lambda_{yy}=\lambda_{ab}^2 
\nonumber \\
\Lambda_{zz}=\lambda_{c}^2. 
\end{eqnarray}
$\lambda_{ab}$ and $\lambda_{c}$ are the magnetic penetration depths
in the ab plane and along the c axis respectively. 
Solving, one obtains for the local transverse field\cite{ehb:95}
\begin{eqnarray}
\label{3ahp}
{\bf h}_{\perp}({\bf r},t)
&=&
\int {d {\bf p} \over (2 \pi)^3} \ e^{+i{\bf p}\cdot{\bf r}}
{\bf h}_{\perp}({\bf p},t)
\nonumber \\
h_{\alpha}({\bf p},t)
&=&
\Phi_o \sum_i \int d {\bf r}_{i\beta}(t) \ 
e^{-i{\bf p}\cdot{\bf r}_i(t)} f_{\alpha \beta}({\bf p})
\end{eqnarray}
where ${\bf r}_i(t)=(x_i(z,t),y_i(z,t),z)$ describes the position of
vortex $i$ at time $t$ and
\begin{eqnarray}
f_{xx}({\bf p})
&=&
{(1 + \lambda_{c}^2 p_x^2 + \lambda_{ab}^2 p_y^2 
                          + \lambda_{ab}^2 p_z^2) 
 \over
 (1 + \lambda_{ab}^2 p^2)
 (1 + \lambda_{c}^2 p_{\perp}^2 + \lambda_{ab}^2 p_z^2)}
\nonumber \\
f_{yy}({\bf p})
&=&
{(1 + \lambda_{ab}^2 p_x^2 + \lambda_{c}^2 p_y^2 
                          + \lambda_{ab}^2 p_z^2) 
 \over
 (1 + \lambda_{ab}^2 p^2)
 (1 + \lambda_{c}^2 p_{\perp}^2 + \lambda_{ab}^2 p_z^2)}
\nonumber \\
f_{zz}({\bf p})
&=&
1
\nonumber \\
f_{xy}({\bf p})=f_{yx}({\bf p})
&=&
{(\lambda_{c}^2 - \lambda_{ab}^2) p_x p_y
 \over
 (1 + \lambda_{ab}^2 p^2)
 (1 + \lambda_{c}^2 p_{\perp}^2 + \lambda_{ab}^2 p_z^2)}
\nonumber \\
f_{xz}({\bf p})=f_{zx}({\bf p})
&=&
0
\nonumber \\
f_{yz}({\bf p})=f_{zy}({\bf p})
&=&
0
\end{eqnarray}

We begin by calculating 
\begin{eqnarray}
K({\bf r},t)
&=&
\langle {\bf h}_{\perp}({\bf r},t) 
        \cdot {\bf h}_{\perp}({\bf r},0) \rangle
\nonumber \\
&=&
\int{d{\bf p} \over (2 \pi)^3}
\int{d{\bf p}^{\prime} \over (2 \pi)^3}
e^{+i({\bf p}+{\bf p}^{\prime})\cdot{\bf r}}
\nonumber \\
& & 
\langle {\bf h}_{\perp}({\bf p},t) 
        \cdot {\bf h}_{\perp}({\bf p}^{\prime},0) \rangle.
\label{krt}
\end{eqnarray}
We are assuming a perfect lattice and hence the position dependence is
completely described by the behavior within the first unit cell.
To obtain this, we average over equivalent lattice sites as well as
over the length of the vortices:
\begin{eqnarray}
K({\vec \rho},t)
&=&
{1 \over N_{\perp}} \sum_{\ell} 
{1 \over L_z} \int dz
K({\bf r},t)
\label{krht1}
\end{eqnarray}
where $N_{\perp}$ is the number of unit cells in a plane perpendicular
to the applied field, $L_z$ is the length along the c axis, and 
${\bf r}={\bf r}_{\ell}^o+{\vec \rho}$.
${\bf r}_{\ell}^o$ is the equilibrium position of vortex $\ell$ and 
$\vec \rho$ is a two dimensional position vector within the first unit
cell. 
Performing this sum and integral we obtain,
\begin{eqnarray}
& & K({\vec \rho},t)
\nonumber \\
& & \hskip 0.1 in
=
{1 \over {\cal A}_{\perp} L_z} \sum_{\bf g}
e^{+i{\bf g}\cdot{\vec \rho}}
\sum_{{\bf g}^{\prime}}
\int_{BZ} {d{\bf k} \over (2 \pi)^3}
\nonumber \\ 
& & \hskip 0.3 in 
\langle
{\bf h}_{\perp}({\bf g}^{\prime}+{\bf k}_{\perp},k_z,t)
\cdot
{\bf h}_{\perp}({\bf g}-{\bf g}^{\prime}-{\bf k}_{\perp},-k_z,0)
\rangle
\label{krht2}
\end{eqnarray}
where ${\cal A}_{\perp}$ is the area of the sample perpendicular to
the c axis,
${\bf g}$ and ${\bf g}^{\prime}$ are reciprocal lattice vectors,
and ${\bf k}$ is restricted to the first Brillouin zone.
The thermal average in the one phonon approximation becomes
\begin{eqnarray}
& &
\langle
{\bf h}_{\perp}({\bf g}^{\prime}+{\bf k}_{\perp},k_z,t)
\cdot
{\bf h}_{\perp}({\bf g}-{\bf g}^{\prime}-{\bf k}_{\perp},-k_z,0)
\rangle
\nonumber \\
& & \hskip 0.1 in
=
B^2 k_z^2 
e^{- {1 \over 2} |{\bf g}^{\prime}+{\bf k}_{\perp}|^2 
     \langle u^2 \rangle}
e^{- {1 \over 2} |{\bf g}-{\bf g}^{\prime}-{\bf k}_{\perp}|^2 
     \langle u^2 \rangle}
\nonumber \\
& & \hskip 0.3 in
f_{\alpha \beta}({\bf g}^{\prime}+{\bf k}_{\perp},k_z)
f_{\alpha \delta}({\bf g}-{\bf g}^{\prime}-{\bf k}_{\perp},-k_z)
\nonumber \\
& & \hskip 0.3 in 
\left[
(\hat{\bf k}_{\perp}\cdot\hat{\bf e}_{\beta})
(\hat{\bf k}_{\perp}\cdot\hat{\bf e}_{\delta})
\langle u_{l}({\bf k},t) u_{l}^*({\bf k},0) \rangle
\right.
\nonumber \\
& & \hskip 0.3 in
\left.
+
(\hat{\bf z}\times\hat{\bf k}_{\perp}\cdot\hat{\bf e}_{\beta})
(\hat{\bf z}\times\hat{\bf k}_{\perp}\cdot\hat{\bf e}_{\delta})
\langle u_{t}({\bf k},t) u_{t}^*({\bf k},0) \rangle
\right]
\label{thavhh}
\end{eqnarray}
where ${\bf u}_i(z,t)$ is the displacement from equilibrium of vortex
$i$ at height $z$ and time $t$
\begin{eqnarray}
{\bf r}_i(t)
&=&
{\bf r}_i^o + {\bf u}_i(z,t)
\end{eqnarray}
and $u_{l}$ and $u_{t}$ are the longitudinal and transverse components
of the lattice distortions
\begin{eqnarray}
{\bf u}({\bf k},t)
&=&
{\Phi_o \over B} \sum_i \int d z \
{\bf u}_i(z,t)e^{-i{\bf k}\cdot{\bf r}_i^o}
\nonumber \\
&=&
u_{l}({\bf k},t) \hat{\bf k}_{\perp}
+ u_{t}({\bf k},t) \hat{\bf z}\times\hat{\bf k}_{\perp}.
\end{eqnarray}
$\alpha, \beta, \delta = x,y$ and repeated subscripts are summed over.
The correlation functions of the distortions, including
$\langle u^2 \rangle$ the mean square vortex displacement,
are obtained using the vortex lattice elastic energy
and the Langevin equation for overdamped motion of the vortices.
\begin{eqnarray}
\label{Hel}
& & {\cal H}_{elastic}
=
{1 \over 2} \int_{BZ} {d {\bf k} \over (2 \pi)^3} \ 
\nonumber \\
& & \hskip 0.6 in
\left\{ 
\epsilon_{l}(k_{\perp},k_z) |u_{l}({\bf k})|^2
+ \epsilon_{t}(k_{\perp},k_z) |u_{t}({\bf k})|^2
\right\} 
\\
& & {\rm with}
\nonumber \\
& & \hskip 0.1 in
\epsilon_{l}(k_{\perp},k_z)
=
c_{11}({\bf k}_{\perp},k_z) k_{\perp}^2 
+ c_{44}({\bf k}_{\perp},k_z) k_z^2
\nonumber \\
& & \hskip 0.1 in
\epsilon_{t}(k_{\perp},k_z)
=
c_{66} k_{\perp}^2 + c_{44}({\bf k}_{\perp},k_z) k_z^2.
\end{eqnarray}
$c_{11}$, $c_{44}$ and $c_{66}$ are the compression, tilt and shear
moduli of the vortex lattice respectively.
The wavevector dependences of the elastic moduli are given in
reference \cite{ehb:95}. 
The equation of motion for the vortices is
\begin{eqnarray}
0
&=&
F_{viscous} + F_{elastic} + F_{noise}
\nonumber \\
&=&
- \eta s {\partial {\bf u}_i \over \partial t}
- {\delta {\cal H}_{elastic} \over \delta {\bf u}_i}
+ \vec \xi
\end{eqnarray}
where $s$ is the layer spacing and $\eta$ is the vortex viscosity
coefficient.
$\eta$ is inversely proportional to the flux flow resistivity.
In YBCO we estimate that it is of order $8 \times 10^{-6}$ g/(cm s)
for fields of order 1 Tesla at 30 Kelvin.\cite{dcm:94}
Using these, 
\begin{eqnarray}
\label{uu}
\langle u_{l}({\bf k},t) u_{l}^*({\bf k},0) \rangle
&=&
{\cal A}_{\perp} L_z \ k_B T \ 
{e^{-\Gamma_{l}(k_{\perp},k_z) t} \over \epsilon_{l}(k_{\perp},k_z)}
\\
{\rm where} & &
\Gamma_{l}(k_{\perp},k_z)
=
{\Phi_o \epsilon_{l}(k_{\perp},k_z)
 \over 2 B \eta} 
\end{eqnarray}
and likewise for the transverse component.
Also,
\begin{eqnarray}
\langle u^2 \rangle
&\approx&
k_B T \left({B \over 4 \pi \Phi_o 
c_{66} c_{44}(k_{BZ},{\pi \over \xi})}\right)^{1/2}
\end{eqnarray}
where $k_{BZ}^2 \equiv {4 \pi B \over \Phi_o}$ gives the size of the
magnetic Brillouin zone. 
Using the parameters for YBCO, which I list in detail below, we find  
$\langle u^2 \rangle$ is of order 10$^2$ \AA$^2$ for fields of order 1
Tesla at 30 Kelvin.
Combining (\ref{uu}), (\ref{thavhh}), and (\ref{krht2}),
and performing the Fourier transform in time from (\ref{krom}),
we obtain
\begin{eqnarray}
K(\vec \rho, \omega)
&=&
{4 B^3 k_B T \eta \over \Phi_o}
\sum_{\bf g} e^{+i{\bf g}\cdot{\vec \rho}}
\sum_{{\bf g}^{\prime}} \int_{BZ} {d{\bf k} \over (2 \pi)^3}
\nonumber \\
& & 
e^{-{1 \over 2}|{\bf g}^{\prime}+{\bf k}_{\perp}|^2
    \langle u^2 \rangle}
e^{-{1 \over 2}|{\bf g}^{\prime}+{\bf k}_{\perp}-{\bf g}|^2
    \langle u^2 \rangle}
\nonumber \\
& & 
k_z^2
f_{\alpha \beta}({\bf g}^{\prime}+{\bf k}_{\perp},k_z)
f_{\alpha \delta}({\bf g}^{\prime}+{\bf k}_{\perp}-{\bf g},k_z)
\nonumber \\
& & 
\left\{
{(\hat {\bf k}_{\perp} \cdot \hat {\bf e}_{\beta})
 (\hat {\bf k}_{\perp} \cdot \hat {\bf e}_{\delta})
 \over
 \left[ \epsilon_{l}({\bf k}_{\perp},k_z) \right]^2
 + \left[ {2 B \eta \omega \over \Phi_o} \right]^2}
\right.
\nonumber \\
& & \hskip 0.1 in
\left.
+
{(\hat {\bf z} \times \hat {\bf k}_{\perp} \cdot \hat {\bf e}_{\beta})
 (\hat {\bf z} \times \hat {\bf k}_{\perp} \cdot \hat {\bf e}_{\delta})
 \over
 \left[ \epsilon_{t}({\bf k}_{\perp},k_z) \right]^2
 + \left[ {2 B \eta \omega \over \Phi_o} \right]^2}
\right\}
\end{eqnarray}
Multiplying by ${1 \over 2}\gamma_n^2$ and setting $\omega=\gamma_n B$
one has a complete expression for the position dependent rate of
relaxation by vortex vibrations.

\subsection{Discussion}
\label{vvdisc}

\begin{center}
\begin{figure*} [t]
\includegraphics*[width=3in]{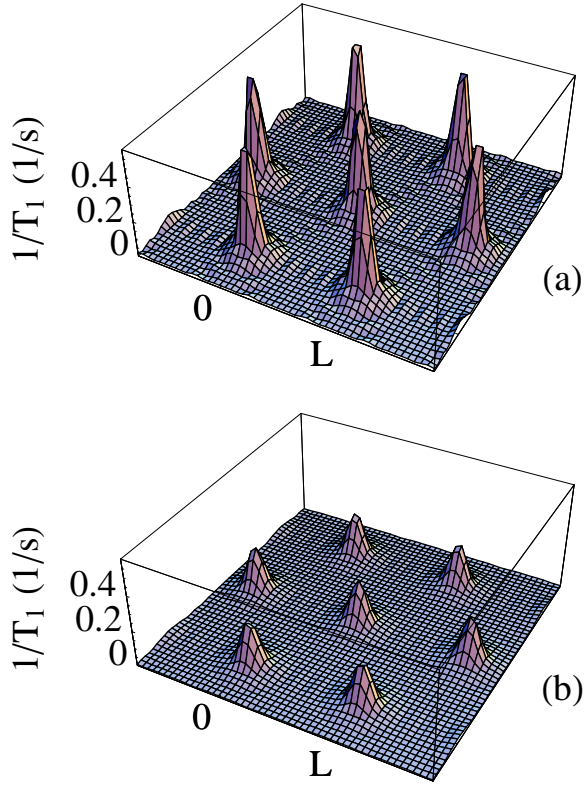}
\caption{The calculated rate of relaxation due to vortex vibrations as
a function of position at (a) 8 kG and 30 Kelvin and (b) 80 kG and 30
K.  $L=(2 \Phi_o / \sqrt{3} B)^{1/2}$ is the spacing between vortices.}
\label{vvwr} 
\end{figure*}
\end{center}

In strongly layered materials, it is possible to make reliable
approximations about which terms in this expression (or its two
dimensional equivalent) will dominate the final result.
However, in the relatively three dimensional material YBCO, such
approximations are not sufficiently accurate to yield useful results.
We have therefore calculated the required sums numerically.
In doing so, we use lattice parameters from YBCO: 
a c-axis lattice constant of 11.68 \AA \ and (neglecting ab anisotropy)
an in-plane lattice constant of 3.855 \AA.
Furthermore, we use an ab-plane correlation length of 16 \AA, an
ab-plane penetration depth of 1500 \AA, and an anisotropy ratio,
$\Gamma$, of 5: 
$\lambda_c=\Gamma \lambda_{ab}$ and
$\xi_c=\xi_{ab}/\Gamma$.
Notice that the momentum sums are cutoff by ${\pi \over \xi}$.
Our results at 30 Kelvin for two different fields are shown in Figure
\ref{vvwr}.
In both cases, over most of the sample the rate of relaxation is
extremely small.
Near the vortices the rate peaks sharply, especially at lower fields.
However, even at the top of these peaks the rate of relaxation is much
slower than what is observed experimentally.

For vortex vibration to be an effective mechanism of relaxation, two
conditions must be met.
First, there must be a significant component of the field transverse
to the direction in which it is applied (${\bf h}_{\perp}$).
Second, the frequency spectrum of the vortex motion must show
significant weight near zero frequency (i.e. at the nuclear Zeeman
splitting). 
The three dimensional vortices in YBCO are stiffer than the stacks of
pancake vortices found in more strongly layered superconductors.
The result is that in YBCO both the magnitude of the transverse field
component and the weight at zero frequency are lower than in two
dimensional materials.
However, relaxation by this mechanism might still be observable were
it not for the presence of faster mechanisms.

\section{Relaxation by electron spin-flip scattering}
\label{esf}

\subsection{Calculating $W^{esf}$}

We now calculate the relaxation rate due to electron spin-flip
scattering, $W^{esf}$.  
Again, we consider only the case of a field applied perpendicular to
the CuO planes.
The nuclear transition rate is the thermal average of the 
transition rate between specific electronic, as well as nuclear,
states:
\begin{eqnarray}
W^{esf}
\equiv
{1 \over 3}
W_{(-{3 \over 2})(-{1 \over 2})}
=
\Big\langle 
{1 \over 3}
W_{(-{3 \over 2},i)(-{1 \over 2},f)} 
\Big\rangle
\end{eqnarray}
where 
$i$ and $f$ are the initial and final many-body states of the electron
system, and 
$\Big\langle \Big\rangle$ indicates a thermal average.
This rate is given by the Golden Rule:
\begin{eqnarray}
W_{(m,i)(n,f)}
&=&
{2 \pi \over \hbar} 
|\langle n, f | V | m, i \rangle |^2 
\delta(E_{m,i} - E_{n,f})
\end{eqnarray}
where $E_{m,i}$ is the total energy of the nuclear state $m$ and
electronic state $i$, and likewise for $E_{n,f}$.
The interaction, $V$, comes from the electron-nuclear hyperfine
coupling. 
Using the model of Shastry\cite{bss:89}, Mila and Rice,\cite{mr:89} 
\begin{eqnarray}
V
&=&
\gamma_e \gamma_n \hbar^2
\left(
A_{\perp} I_{+}(0) S_{-}(0)
+
\sum_{n} B_{\perp} I_{+}(0) S_{-}({\bf r}_{n})
\right)
\end{eqnarray}
where ${\bf r}_n$ are the positions of the four nearest neighbor
coppers.
The values of $A_{\perp}$ and $B_{\perp}$ have been estimated by comparison 
with Knight shift data.\cite{bp:95,zbp:96}
\begin{eqnarray}
\gamma_e \gamma_n \hbar^2 B_{\perp}
&\approx&
3.06 \times 10^{-19} \ {\rm erg}
\nonumber \\
{A_{\perp} \over B_{\perp}}
&\approx&
0.8
\end{eqnarray}
We have
\begin{eqnarray}
& & W^{esf}
\nonumber \\
& & \hskip 0.1 in
=
{1 \over 3}
{2 \pi \over \hbar}
\left( \gamma_e \gamma_n \hbar^2 \right)^2
|\langle -{1 \over 2} | I_{+}(0) | -{3 \over 2} \rangle |^2
\nonumber \\
& & \hskip 0.2 in
\Big\langle
|\langle f | A_{\perp} S_{-}(0) 
             + \sum_{n.n.} B_{\perp} S_{-}({\bf r}_n) | i \rangle |^2
\delta(E_i - E_f)
\Big\rangle
\label{wesf3}
\end{eqnarray}
Note that we may neglect the {\it nuclear} Zeeman splitting in the
energy $\delta$-function because it is so much smaller than the
electron's. 
The nuclear matrix element just gives a factor of three.

In order to calculate the matrix elements of the operator
\begin{eqnarray}
S_{-}({\bf r})
&=&
\Psi_{\downarrow}^{\dag}({\bf r})
\Psi_{\uparrow}({\bf r}),
\end{eqnarray}
we write the electron operators, $\Psi$, in terms of Bogoliubov
quasiparticle operators:
\begin{eqnarray}
\Psi_{\uparrow}({\bf r})
&=&
{1 \over \sqrt{N_{\perp}}} \sum_{\bf k}
\left(
\gamma_{{\bf k}\uparrow} u_{{\bf k}\uparrow}({\bf r})
-
\gamma_{{\bf k}\downarrow}^{\dag} v_{{\bf k}\downarrow}^*({\bf r})
\right)
\nonumber \\
\Psi_{\downarrow}({\bf r})
&=&
{1 \over \sqrt{N_{\perp}}} \sum_{\bf k}
\left(
\gamma_{{\bf k}\downarrow} u_{{\bf k}\downarrow}({\bf r})
+
\gamma_{{\bf k}\uparrow}^{\dag} v_{{\bf k}\uparrow}^*({\bf r})
\right)
\end{eqnarray}
where $N_{\perp}$ is the number of nuclear sites in a plane
perpendicular to the c axis.
The values of $u_{{\bf k}\alpha}$ and $v_{{\bf k}\alpha}$ 
are obtained by solving the Bogoliubov-deGennes equations.
In applying the magnetic field, $H_o$, we make the simplifying
assumption that the spatial variation which arises from the formation
of vortices is over sufficiently long length scales that at each
point, ${\bf r}$, we can simply include a uniform superfluid flow,
${\bf v}_s({\bf r})$, and Zeeman splitting. 
We obtain
\begin{eqnarray}
u_{{\bf k}\alpha}({\bf r})
&=&
{e^{+i({\bf k}+{\bf q})\cdot{\bf r}} \over \sqrt{2}}
\sqrt{
1 +
{\xi_{\bf k} \over 
 \pm \sqrt{\xi_{\bf k}^2 + \Delta_{\bf k}^2}}
}
\nonumber \\
v_{{\bf k}\alpha}({\bf r})
&=&
\pm ({\rm sgn}\Delta_{\bf k})
{e^{+i({\bf k}-{\bf q})\cdot{\bf r}} \over \sqrt{2}}
\sqrt{
1 -
{\xi_{\bf k} \over 
 \pm \sqrt{\xi_{\bf k}^2 + \Delta_{\bf k}^2}}
}
\label{uandv}
\end{eqnarray}
where $2 \hbar {\bf q}$ is the center-of-mass momentum of the Cooper
pairs (${\bf q}=- k_F {\bf v}_s / 4 v_F$).
The sign of the square root in each case is determined by the
requirement that $\epsilon_{{\bf k}\alpha}$ be positive:
\begin{eqnarray}
\epsilon_{{\bf k}\alpha}
&=&
- ({\rm sgn}\alpha){1 \over 2} \gamma_e \hbar H_o
+ \hbar {\bf k} \cdot {\bf v}_s
\pm \sqrt{\xi_{\bf k}^2 + \Delta_{\bf k}^2}
\label{epska}
\end{eqnarray}
where $\xi_{\bf k}$ are the normal state excitation energies,
and $\Delta$ is the superconducting gap.
\cite{note}

We can now calculate the thermal average in equation (\ref{wesf3}).
The result is a sum on wavevectors ${\bf k}$ and ${\bf k}^{\prime}$  
over terms containing one
of the following three $\delta$-functions:
$\delta(\epsilon_{{\bf k}\uparrow}
        -\epsilon_{{\bf k}^{\prime}\downarrow})$,
$\delta(\epsilon_{{\bf k}\uparrow}
        +\epsilon_{{\bf k}^{\prime}\uparrow})$, or
$\delta(\epsilon_{{\bf k}\downarrow}
        +\epsilon_{{\bf k}^{\prime}\downarrow})$.
Because the quasiparticle excitation energies 
$\epsilon_{{\bf k}\alpha}$ must, by definition, be positive, the
latter two $\delta$-functions restrict the resulting energy integrals
to a region with a width equal to the {\it nuclear} Zeeman splitting.
The result is vanishingly small.
We are left, therefore, with the following expression:
\begin{eqnarray}
& &
\Big\langle
|\langle f | A_{\perp} S_{-}(0) 
             + \sum_{n.n.} B_{\perp} S_{-}({\bf r}_n) | i \rangle |^2
\delta(E_i - E_f)
\Big\rangle
\nonumber \\
& & \hskip 0.1 in
=
{1 \over N_{\perp}^2} 
\sum_{{\bf k}} \sum_{{\bf k}^{\prime}}
\nonumber \\
& & \hskip 0.3 in
\left[
A_{\perp} + 2 B_{\perp} 
\left( \cos a(k_x-k_x^{\prime}) + \cos a(k_y-k_y^{\prime}) \right)
\right]^2
\nonumber \\
& & \hskip 0.3 in
f(\epsilon_{{\bf k}\uparrow}) 
[1-f(\epsilon_{{\bf k}^{\prime}\downarrow})]
\delta(\epsilon_{{\bf k}\uparrow}
       -\epsilon_{{\bf k}^{\prime}\downarrow})
\nonumber \\
& & \hskip 0.3 in
\left(
u_{{\bf k}\uparrow} u_{{\bf k}^{\prime}\downarrow}
+
v_{{\bf k}\uparrow} v_{{\bf k}^{\prime}\downarrow}
\right)^2
\end{eqnarray}
To evaluate this expression, we begin by approximating the sums by 
integrals.  Furthermore, because of the Fermi functions, the important
regions in ${\bf k}$ space will be near the intersections of each of
the four nodes with the Fermi surface, we convert the integral over
all ${\bf k}$ space to a sum over the four nodes of an integral
centered at an intersection: 
\begin{eqnarray}
{1 \over N_{\perp}}
\sum_{\bf k}
=
{{\cal A}_{\perp} \over N_{\perp}} \int {d \bf k \over (2 \pi)^2}
=
a^2 \sum_{n} \int {d k_{n \perp} d k_{n \|} \over (2 \pi)^2}
\end{eqnarray}
where 
${\cal A}_{\perp}$ is the area of the sample perpendicular to the c axis,
and $k_{n \perp}$ and $k_{n \|}$ are tangent to and normal to the
Fermi surface at node $n$, respectively.
Because $k_{n \perp}$ and $k_{n \|}$ are both much less than $k_F$, 
we neglect them in calculating the $\cos[a(k-k^{\prime})]$ factors.
In this case,
\begin{eqnarray}
& &
\cos[a(k_{nx}-k_{n^{\prime}x})]
+
\cos[a(k_{ny}-k_{n^{\prime}y})]
\nonumber \\
& & \hskip 0.5 in
=
2 \cos (n-n^{\prime}){\pi \over 2}
\end{eqnarray}
where $n,n^{\prime}=1,2,3,4$ label the four nodes.

We approximate the values of $\xi_{\bf k}$ and $\Delta_{\bf k}$ as
follows:
\begin{eqnarray}
\xi_{\bf k}
&\approx&
\hbar v_F k_{\|}, \quad {\rm and}
\nonumber \\
\Delta_{\bf k}
&\approx&
\hbar v_1 k_{\perp}
\end{eqnarray}
where $v_F$ is the Fermi velocity, 
and $v_1$ is the slope of the gap near the node.
This allows us to make the following transformation from Cartesian to
polar coordinates:
\begin{eqnarray}
\sqrt{\xi_{\bf k}^2 + \Delta_{\bf k}^2}
\approx
\sqrt{(\hbar v_F k_{\|})^2 + (\hbar v_1 k_{\perp})^2}
&=&
E
\nonumber \\
{\xi_{\bf k} \over \Delta_{\bf k}}
\approx
\tan^{-1} \left({v_F k_{\|} \over v_1 k_{\perp}}\right)
&=&
\theta
\end{eqnarray}
The Jacobian for this transformation is $E/(\hbar^2 v_F v_1)$.
Thus
\begin{eqnarray}
& & {1 \over N_{\perp}^2}
\sum_{\bf k} \sum_{{\bf k}^{\prime}}
\rightarrow
\left( {a^2 \over (2 \pi)^2} \right)^2
{1 \over \left( \hbar^2 v_F v_1 \right)^2}
\nonumber \\
& & \hskip 1 in
\sum_n \sum_{n^{\prime}}
\int E dE d\theta \int E^{\prime} dE^{\prime} d\theta^{\prime}
\end{eqnarray}
Finally, we write $E$ and $E^{\prime}$ in terms of the energies of the
initial and final states of the scattered electron as given by
equation (\ref{epska}):
\begin{eqnarray}
\epsilon
&=&
D_{n\uparrow} + \sigma E
\nonumber \\
\epsilon^{\prime}
&=&
D_{n^{\prime}\downarrow} + \sigma^{\prime} E^{\prime}
\end{eqnarray}
where
$\sigma=-1$ for $0 < \epsilon \leq {\rm max}[0,D_{n\uparrow}]$
and $\sigma=+1$ for $\epsilon > {\rm max}[0,D_{n\uparrow}]$
and likewise for $\epsilon^{\prime}$;
and where
\begin{eqnarray}
& & 
D_{n\uparrow}(r,\phi)
\nonumber \\
& & \hskip 0.1 in
=
- 
{1 \over 2} \gamma_e \hbar H_o
+ 
\hbar k_F v_s(r) \cos 
\left( \phi + (2n-1){\pi \over 4} \right)
\nonumber \\
& &
D_{n^{\prime}\downarrow}(r,\phi)
\nonumber \\
& & \hskip 0.1 in
=
+ 
{1 \over 2} \gamma_e \hbar H_o
+ 
\hbar k_F v_s(r) \cos 
\left( \phi + (2n^{\prime}-1){\pi \over 4} \right).
\label{deltas}
\end{eqnarray}
In these expressions, we have made two simplifying assumptions.
First, as mentioned above, $k_{n\|}$ and $k_{n\perp}$ are much less 
than $k_F$ and hence $|{\bf k}| \approx |{\bf k}_n| = k_F$.
Second, we have simplified the geometry of the system by treating the
lattice as a collection of identical cylindrical cells each centered
on a single vortex.  
We neglect nonuniformity along the length of the vortices and allow 
the magnitude of the supercurrent velocity to vary only as a function
of distance from the vortex core. 
Due to the asymmetry of the superconducting gap, however, some angular
dependence remains.

The angles $\theta$ and $\theta^{\prime}$ now enter only in the
coherence factor:  
\begin{eqnarray}
& &
\left(
u_{{\bf k}\uparrow} u_{{\bf k}^{\prime}\downarrow}
+
v_{{\bf k}\uparrow} v_{{\bf k}^{\prime}\downarrow}
\right)^2
\nonumber \\
& & \hskip 0.1 in
=
{1 \over 2}
\left[
1 + \sigma \sigma^{\prime}
\left(
\sin \theta \sin \theta^{\prime}
\right. \right.
\nonumber \\
& & \hskip 1 in
\left. \left.
+
({\rm sgn}\Delta)({\rm sgn}\Delta^{\prime})
|\cos \theta| |\cos \theta^{\prime}|
\right)
\right]
\end{eqnarray}
When this is integrated over $\theta$ and $\theta^{\prime}$ one
obtains ${1 \over 2}(2 \pi)^2$ for all possible combinations of
signs. 

Combining all of these points, we have
\begin{eqnarray}
& &
W^{esf}
\nonumber \\
& & \hskip 0.1 in
=
{2 \pi \over \hbar}
(\gamma_e \gamma_n \hbar^2)^2 
\left( {a^2 \over (2 \pi)^2} \right)^2
{1 \over \left( \hbar^2 v_F v_1 \right)^2}
{1 \over 2}(2 \pi)^2
\nonumber \\
& & \hskip 0.3 in
\sum_{n} \sum_{n^{\prime}}
\Biggl[
A_{\perp}
+
4 B_{\perp} \cos (n-n^{\prime}){\pi \over 2} 
\Biggr]^2
\nonumber \\
& & \hskip 0.3 in
\sum_{\sigma,\sigma^{\prime}=\pm 1}
\int_0^{\infty} E \ dE 
\int_0^{\infty} E^{\prime} \ dE^{\prime}
\nonumber \\
& & \hskip 0.3 in
f(D_{n\uparrow} + \sigma E)
[1-f(D_{n^{\prime}\downarrow} + \sigma^{\prime} E^{\prime})] 
\nonumber \\
& & \hskip 0.3 in
\delta\left( (D_{n\uparrow} + \sigma E) 
 - (D_{n^{\prime}\downarrow} + \sigma^{\prime} E^{\prime}) \right)
\nonumber \\
& & \hskip 0.3 in
\Theta(D_{n\uparrow} + \sigma E) 
\Theta(D_{n^{\prime}\downarrow} + \sigma^{\prime} E^{\prime})
\end{eqnarray}
where $\Theta$ is the Heaviside function.
The form of this is simply a sum over initial and final electron
states of the product of
(i) a factor arising from the nonlocal form of the interaction,
(ii) the density of initial states, 
(iii) the density of final states,
(iv) the probability that the initial state is occupied,
(v) the probability that the final state is empty,
and finally (vi) an energy conserving $\delta$-function.
Evaluating one of the integrals, we obtain
\begin{eqnarray}
& &
W^{esf}(r,\phi)
\nonumber \\
& & \hskip 0.1 in
=
{1 \over 16 \pi \hbar}
\left(
{\gamma_e \gamma_n \hbar^2 B_{\perp} a^2
 \over \hbar^2 v_F v_1}
\right)^2
(2 k_B T)^3
\nonumber \\
& & \hskip 0.3 in
\sum_n \sum_{n^{\prime}}
\Biggl[
{A_{\perp} \over B_{\perp}}
+
4 \cos (n-n^{\prime}){\pi \over 2} 
\Biggr]^2
\nonumber \\
& & \hskip 0.2 in
\int_{0}^{\infty}
\Big{|} x - {D_{n\uparrow}(r,\phi) \over 2 k_B T} \Big{|} \ 
\Big{|} x - {D_{n^{\prime}\downarrow}(r,\phi) \over 2 k_B T} \Big{|} \ 
{\rm sech}^2 x \ dx
\end{eqnarray}

\subsection{Discussion}

In the absence of a magnetic field, there is no Zeeman splitting, no
Doppler shifting and all positions are identical.  
The integral above is then trivial, and 
the rate of relaxation at all points is
\begin{eqnarray}
W_{H_o=0}^{esf}
&=&
{1 \over 16 \pi \hbar}
\left(
{\gamma_e \gamma_n \hbar^2 B_{\perp} a^2
 \over \hbar^2 v_F v_1}
\right)^2
(2 k_B T)^3
{\pi^2 \over 12}
\nonumber \\
& &
\sum_n \sum_{n^{\prime}}
\Biggl[
{A_{\perp} \over B_{\perp}}
+
4 \cos (n-n^{\prime}){\pi \over 2} 
\Biggr]^2
\nonumber \\
&\approx& 
1.4 \ {\rm s}^{-1}
\qquad {\rm at \ 30 \ Kelvin}
\label{w1min}
\end{eqnarray}
where we have used $a \approx 3.855$ \AA, 
$v_F \approx 1.3 \times 10^{7} \ {\rm cm/s}$,
and $v_1 \approx v_F/(14)$.\cite{mc:99}
The probability distribution, $P(W_1)$ from (\ref{pofw1}), is just a 
$\delta$-function at this value.
This fits the NQR data very well.\cite{data}  

In the presence of a magnetic field, it is no longer possible to
express the rate at an arbitrary position in closed form.
This is due to terms of the form
\begin{eqnarray}
\int_{a}^{b} x^2 \ {\rm sech}^2 x \ dx
\end{eqnarray}
where $a$ is not zero and $b$ is not infinity.
It is useful, however to look at two limiting cases.

First, far from any vortex core, the Doppler shift is negligibly small
relative to the thermal energy.
If the Zeeman splitting is also negligible, the rate at these
positions is the same as in zero applied field:
\begin{eqnarray}
W_{min}^{esf} &=& W_{H_o=0}^{esf}.
\end{eqnarray}
At 30 Kelvin, the Zeeman splitting is much less than the thermal
energy up to fields of 10 Tesla, and may be safely neglected. 
At lower temperatures, however, the Zeeman splitting may become
significant.
As the Zeeman splitting is increased from zero, the effect is
initially to {\it lower} the relaxation rate below the zero field
value.
Essentially this is because, while in zero field the product of the
initial and final densities of states has the form $E^2$, with Zeeman
splitting it has the form $|E-Z|\ |E+Z|=|E^2-Z^2|$, where 
$Z \equiv ({1 \over 2}\hbar \gamma_e H_o)/(2 k_B T)$.
As the Zeeman splitting is further increased, such that the zero in
the product of density of states moves outside the range of the Fermi
functions, the rate rises again.
The minimum rate occurs at a Zeeman splitting of roughly half the
thermal energy and is about 20\% below the zero field rate.
The rate climbs above the zero field rate when the Zeeman splitting
reaches roughly 85\% of the thermal energy. 

It is also possible to make a good estimate of the rate of
relaxation near the vortices where the Doppler shift is much larger
than the thermal energy.
Here we may safely neglect the Zeeman splitting.
Because the Doppler shift depends on the angle between the superflow
and the node directions, the rate of relaxation also depends on this
angle (although the magnitude of the variation is small, only about 
10\%).
The value of this angle at which the rate is maximum varies between
$\pi/4$ at low temperatures and 0 at high temperatures, with a
crossover around 20 K.
At an angle of $\pi/4$, the Doppler shifts associated with two of the
nodes are $+D_{max}/\sqrt{2}$ and with the other two 
$-D_{max}/\sqrt{2}$.
If $D_{max}/(2 \sqrt{2} k_B T)>>1$, we may neglect integrals from 
$D_{max}/(2 \sqrt{2} k_B T)$ to infinity and keep only those from
zero to $D_{max}/(2 \sqrt{2} k_B T)$.
We then obtain
\begin{eqnarray}
W_{max}^{esf}
&\approx&
{1 \over 16 \pi \hbar}
\left(
{\gamma_e \gamma_n \hbar^2 B_{\perp} a^2
 \over \hbar^2 v_F v_1}
\right)^2
(64)
\nonumber \\
& & 
\Biggl\{
(2 k_B T) D_{max}^2 
\left(1+{1 \over 8}\left({A_{\perp} \over B_{\perp}}\right)^2\right)
\nonumber \\ 
& & \hskip 0.1 in
+
(2 k_B T)^3 \left( {\pi^2 \over 12}{A_{\perp} \over B_{\perp}}\right)
\Biggr\}
\nonumber \\
&\approx&
10 \ {\rm s}^{-1}
\qquad {\rm at \ 30 \ Kelvin}
\label{w1max}
\end{eqnarray}
Here, in addition to the parameters used in the zero field estimate, 
I have assumed that the maximum Doppler shift is roughly given by
the maximum value of the superconducting gap which is approximately
200 Kelvin.

\begin{center}
\begin{figure*} [t]
\includegraphics*[width=3in]{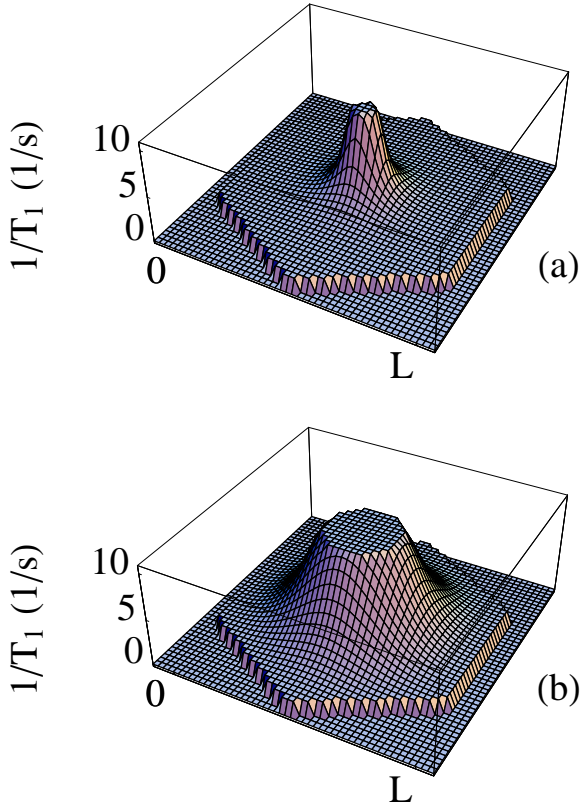}
\caption{The calculated rate of relaxation due to electron spin-flip
scattering as a function of position at (a) 8 kG and 30 K and (b) at
80 kG and 30 K.
$L=(2 \Phi_o / \sqrt{3} B)^{1/2}$ is the spacing between vortices.}
\label{esfwr}
\end{figure*}
\end{center}
Figure \ref{esfwr} shows the rate of relaxation as a function of
position in one vortex unit cell for two different values of applied
field. 
To construct these sketches and the remaining figures, we have
calculated the position dependence of the superfluid velocity from
\begin{eqnarray}
{\bf v}_s({\bf r})
&\propto&
\sum_{\bf Q}
{(\hat {\bf z} \times {\bf Q}) \sin{\bf Q}\cdot{\bf r}
 e^{-Q^2 xi^2/2}
 \over 1+\lambda^2 Q^2}
\end{eqnarray}
where the magnetic penetration depth, $\lambda$, is 1600 \AA,
the superconducting correlation length, $\xi$, is 16 \AA, 
and ${\bf Q}$ is summed over the reciprocal lattice vectors of the
vortex lattice.
The Gaussian factor provides a smooth short distance cutoff.
We scale the magnitude such that the maximum value of the Doppler
shift, reached at roughly 1.5$\xi$ from the core, 
is 200 K.
Inside this radius, we set the rate of relaxation equal to its maximum
value, as if there were a region of normal fluid in the core.
This choice may be controversial; however, it has little
influence on the final form of the magnetization recovery as the cores
account for only a small fraction of the sample area.
What we see in Figure \ref{esfwr} is that most of the area of the
sample is relaxed at or near the thermal relaxation rate.
The rate then rises sharply towards the vortex cores.

\begin{center}
\begin{figure*} [t]
\includegraphics[width=3in,angle=270]{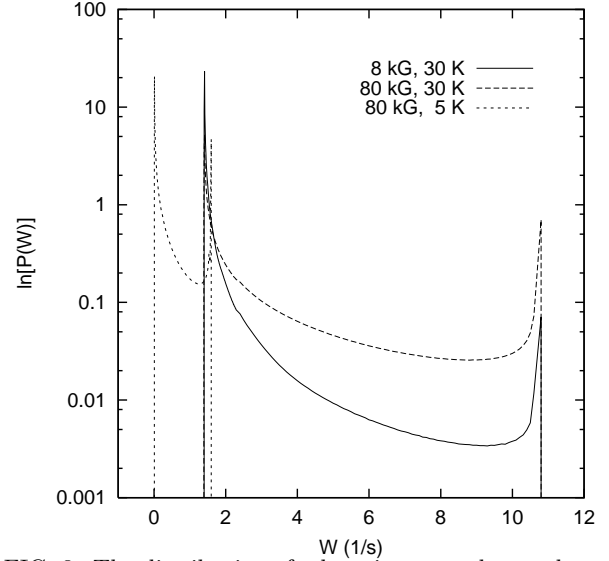}
\caption{The distribution of relaxation rates due to electron
spin-flip scattering.}
\label{esfpw}
\end{figure*}
\end{center}

The quantitative information in Figure~\ref{esfwr} is shown more
explicitly in Figure~\ref{esfpw}.  
This shows the rate distribution--the probability that a spin will be
relaxed at a given rate--at both field values shown in Figure 2 as
well as a lower temperature value for comparison.
In the absence of an applied field, the distribution would be a
$\delta$-function at the thermal rate of relaxation
(Eqn.~\ref{w1min}). 
In the presence of an applied field, a sharp peak remains at the
thermal rate, but the distribution is spread between this and the
large Doppler shift limit (Eqn.~\ref{w1max}).
At low fields, near $H_{c1}$, the high rate tale is completely
negligible.
At higher fields, more weight is shifted to higher rates;
however, even at fields of 10 Tesla (at 30 Kelvin) one must use
a log scale to see the shape of the distribution.
Because the lower end of the distribution is proportional to $T^3$
while the upper end is roughly proportional to $T \Delta_{max}^2$,
for the temperature range of interest (well below $\Delta_{max}$) 
the width of the distribution increases as the temperature is
increased. 

\begin{center}
\begin{figure*} [t]
\includegraphics[width=3in,angle=270]{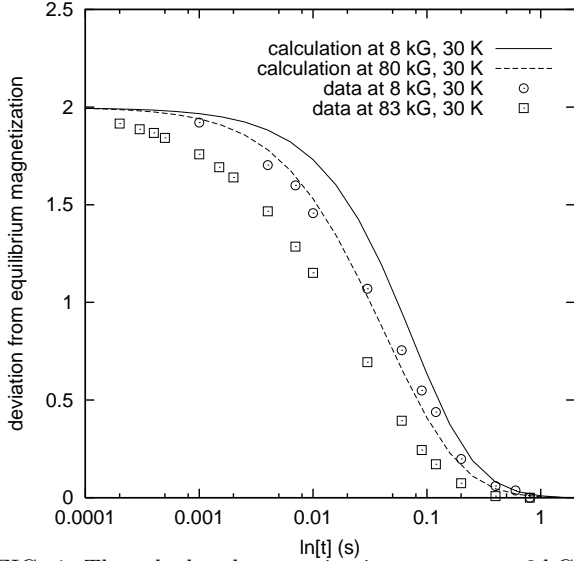}
\caption{The calculated magnetization recovery at 8 kG and 80 kG and
data taken at 8 kG and 83 kG, all at 30 K.}
\label{esfmt}
\end{figure*}
\end{center}
Figure \ref{esfmt} shows the magnetization recovery curves
corresponding to the rate distributions from Fig.~\ref{esfwr} and
compares them with experimental data from Reference~\cite{data}.
These are upper satellite inversion recovery measurements on the
planar $^{63}$Cu nuclei of a c-axis aligned powder of
YBa$_2$Cu$_3$O$_{7-\delta}$ with the field applied perpendicular to
the planes.
The equilibrium magnetization and the inversion fraction have been
estimated from the raw data to allow presentation in this format.
These electron spin-flip scattering results come much closer to
describing the data than did the vortex vibration work.
However, 
the data exhibit faster relaxation than our calculations predict, 
a broader distribution of relaxation rates (as seen in their more
gradual decline) and a stronger field dependence.
	
\begin{center}
\begin{figure*} [t]
\includegraphics*[width=3in]{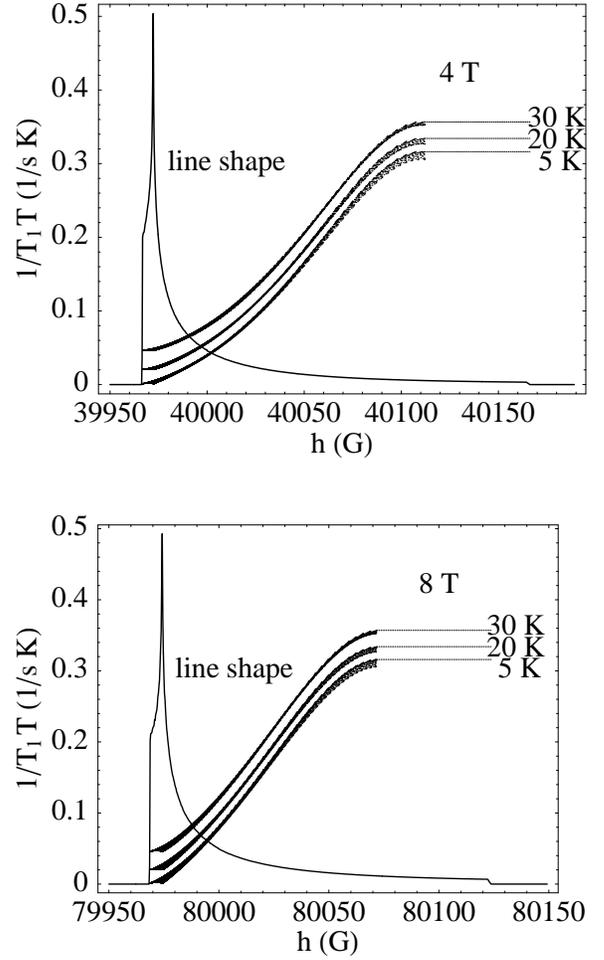}
\caption{The rate of relaxation, divided by temperature, as a function
of position in the resonance line at 5, 10 and 30 K in 4 and 8 T
fields.} 
\label{esfwh}
\end{figure*}
\end{center}
Finally, in addition to these measurements of the relaxation of the
full resonance line, recent experiments study the relaxation as a
function of frequency within the line.\cite{new,O17}
The resonance line in the presence of a vortex lattice has a
characteristic shape, shown in Figure \ref{esfwh}.
$T_1$ data taken as a function of frequency within this line would
allow direct observation of the position dependence arising from
Doppler shifted quasiparticle states.
Figure \ref{esfwh} presents the rate of relaxation divided by
temperature expected as a function of position within the resonance
line at two different fields and three different temperatures.
At each field there is a narrow distribution of relaxation rates
due to the fact that the angular variation of the magnetic field
around a vortex is different from the angular variation of the
relaxation rate (which is also temperature dependent).

\section{Conclusion}
\label{concl}

In this paper, we have examined the two most likely sources of
position dependent spin-lattice relaxation in the mixed state of YBCO.
First, assuming overdamped harmonic vibrations of the vortex lattice
with parameters appropriate to YBCO, we find that relaxation by vortex
vibrations, although strongly position dependent, is too slow to play
a significant role in the observed magnetization recovery.
Second, we have studied electron spin-flip scattering within a
framework of noninteracting quasiparticles with energies Doppler
shifted by the supercurrents circulating around vortices.
Again our results are strongly position dependent.  
In addition, the rates we calculate are similar to those seen in
experiments.
However, significant discrepancies remain between our calculations and
what is observed:
our calculations predict slower rates, a narrower distribution, and
less field dependence than are seen in the experiments.

Given the emphasis we have placed on the nonuniformity of relaxation
rates, it might appear that nuclear spin diffusion should also be
considered.  This is especially so given the strong {\it indirect}
nuclear dipole coupling through electron spins in the cuprates. 
If spin diffusion were to take place, it would smooth out the spatial
variation in relaxation rates and hence widen the gap between our
results and the experiments.  
However, spin diffusion is very strongly suppressed in the nonuniform
magnetic field of the vortex lattice, especially so near the vortex
cores where the field gradient (and also the $T_1$ variation) is
greatest.\cite{rw:98}

The fact that this theory which involves a number of simplifying
assumptions and experimental data with attendant complications do not
agree completely is not so surprising.
Regarding the theory,
the fastest relaxation rates come from regions near the vortex cores.
It is precisely in this region where our quasiclassical approximation
and our assumption of linear gap variation break down.
Hence, our theory is least reliable in its description of the
fastest rates.
Recent numerical work\cite{tim:99} may help clarify the behavior in
this region.
Regarding the experiments,
impurities, inhomogeneities and practical difficulties may influence
the data:
Fluctuating moments of magnetic impurities might increase the rate of
relaxation of nearby nuclear sites.
An inhomogeneous superfluid density due to defects in the grains could
complicate the position dependence of the superfluid velocity and
possibly increase the area of large Doppler shifting.
Failure to invert spins across the full vortex-lattice field spectrum,
particularly tricky in the presence of a distribution of quadrupolar
splittings and grain misalignments, might also influence the
data; although the effect in this case would be to narrow the observed
rate distribution.

Doing experiments on clean single crystals would of course settle some
of these issues.
However, returning to the issue of our approximations, 
what experiments would focus most clearly on the features which the
present calculation captures? 
The goal is to work at a temperature and field which maximize the
percentage of spins, $\cal P$, for which the magnitude of the Doppler
shift is (i) larger than the thermal energy (and the Zeeman splitting)
but (ii) much less than the maximum gap.
The first condition is necessary for the effects of the Doppler
shifting to be observable.
The second condition ensures that our approximations are valid.
Clearly going to low temperatures helps to satisfy the first
condition.
This is limited by the fact that the Zeeman splitting will eventually
become larger than the thermal energy.
However, for temperatures above 1 Kelvin, the fields which maximize
$\cal P$ are too small for the Zeeman splitting to be an issue.
Practical issues, such as minimizing impurity effects, will probably
place greater restrictions on the choice of temperature.  
The following results are based on an approximate form of the 
superfluid velocity in which $v_s(r) \propto (1/r)-(1/L)$ where
$r$ is the distance from the vortex core.
If we set the upper bound on the Doppler shift at $\Delta_{max}/5$, at
2 Kelvin a field of 2 Tesla will maximize $\cal P$ at 70\% (with
roughly 10\% of the spins outside the range of our approximations and
20\% relaxing at the thermal rate).
At 5 Kelvin, a field of 4 Tesla maximizes $\cal P$ at 55\% (with
roughly 20\% of the spins outside the range of our approximations and
25\% relaxing at the thermal rate).

Finding clear confirmation of the presence of Doppler shifted
quasiparticles would be a significant achievement in itself.
In addition, $T_1$ measurements might provide insight into the nature
of the quasiparticle interactions.
In a parallel calculation of the electron spin-flip scattering nuclear 
spin-lattice relaxation rate within the framework of antiferromagnetic
spin fluctuations interacting via short-lived quasiparticles,\cite{mw:99}
many of the qualitative features of the relaxation are similar to the
noninteracting quasiparticle model presented here.
However, some features unique to each model exist, and if observed 
could clarify which model is most appropriate to YBCO.

R.W. gratefully acknowledges valuable discussions with 
C.P. Slichter, D.K. Morr, 
N. Curro, R. Corey, C. Milling, 
A.J. Leggett
and J.S. Preston.
This work has been supported by the Natural Sciences and Engineering
Research Council of Canada and by the Aspen Center for Physics.

%\bibliography{nmr}

\begin{thebibliography}{10}
\bibitem{note}
It is perhaps worth emphasizing that we have chosen to define our quasiparticle
  states such that all their energies are positive, i.e. there are no occupied
  quasiparticle states in the ground state. This is mathematically equivalent
  to choosing only the positive sign of the square root in the above
  expression, in which case some quasiparticle states have negative energy and
  are occupied in the ground state. Both approaches produce the same final
  result, although the details of the intervening calculation differ.
\end{thebibliography}

\end{document}